\documentclass[12pt]{iopart}
\usepackage{iopams}
\usepackage{setstack}
\usepackage[latin1]{inputenc}
\usepackage[T1]{fontenc}
\usepackage{graphicx}
\usepackage{graphics}
\usepackage{pstricks}
\usepackage{epsfig}
\usepackage{subfigure}

\begin{document}
\def\theequation{\arabic{section}.\arabic{equation}}
\title{The Cauchy problem of $f(R)$ gravity}
\author{Nicolas Lanahan-Tremblay and Valerio 
Faraoni}
\address{Physics Department, Bishop's University, Sherbrooke,
Qu\`{e}bec, Canada J1M~0C8}
\ead{ntremblay05@ubishops.ca, vfaraoni@ubishops.ca}
\begin{abstract}
The   initial value problem of metric and  Palatini $f(R)$ 
gravity 
is studied by using the dynamical 
equivalence between these theories and Brans-Dicke gravity. The Cauchy problem is well-formulated for 
metric $f(R) $ gravity in the presence of matter and well-posed 
in vacuo. For  Palatini $f(R)$ gravity, 
instead, the Cauchy problem is not well-formulated. 
 \end{abstract} \vspace{1truecm}
\maketitle
\def\theequation{\arabic{section}.\arabic{equation}}

\section{Introduction}
\setcounter{equation}{0}
\setcounter{page}{2}

The study of type 1a supernovae led to the 1998 discovery of the acceleration  
of the 
cosmic expansion  \cite{SN}, which has prompted theoretical  
physicists to look for an 
explanation. Many  models of this accelerated dynamics postulate the existence of dark 
energy, a diffuse mysterious form of 
energy with exotic equation of state $P \approx -\rho$ which amounts to  
70\% of the critical energy density, with the remaining 30\% comprised of dark 
 and ordinary matter. As an alternative to exotic dark energy models, it  has been 
proposed 
that infrared modifications of gravity at the largest scales could  be the explanation 
of the cosmic acceleration  \cite{MG1}. The idea is  either to 
consider the Einstein-Hilbert  action\footnote{Here 
$\kappa \equiv 8\pi 
G$, $ R$ is the Ricci curvature, $\Lambda$ is the cosmological constant, and $g$ is the 
determinant of the metric tensor 
$g_{ab}$. We follow the notations of Ref.~\cite{Wald}.} with 
cosmological constant 
\begin{equation}
S_{EH} = \frac{1}{2\kappa}\int d^{4}x\sqrt{-g}\left(R-2\Lambda\right) + 
 S^{(matter)}
\end{equation}
or some form of dark energy mimicking $\Lambda$, or to change 
gravity as described by the alternative action 
\begin{equation}
\label{2}
S = \frac{1}{2\kappa}\int d^{4}x\sqrt{-g} \, f(R) + S^{(matter)} \;,
\end{equation}
where $f(R)$ is a non-linear function of $R$. This ``$f(R)$'' or 
``modified'' gravity has  a long history (see  
Ref.~\cite{Schmidt} for an 
historical 
perspective): it was originally conceived as a 
mathematical alternative to Einstein's theory  
\cite{EddingtonBuch};  
quadratic 
corrections to the 
Einstein-Hilbert action were discovered to be necessary for the 
renormalization of general relativity  \cite{renorma}, and then 
used to 
generate  
inflation in the early universe without  scalar fields 
\cite{Starobinsky}; see Ref.~\cite{NOD} for a more modern 
perspective.

Modified gravity comes in three possible forms: the  first is 
metric  $f(R)$  
gravity, 
in which the metric is the only independent variable and the  variation of 
the action 
(\ref{2}) with respect to $g^{ab}$ yields the fourth-order field 
equations
\begin{equation}\label{fieldeqs}
f^\prime(R)R_{ab}-\frac{f(R)}{2}g_{ab} = \nabla_{a}\nabla_{b}f^\prime(R
)-g_{ab}\square f^\prime(R)+\kappa \, T_{ab}^{(m)} \;.
\end{equation}
The second possibility, named ``Palatini $f(R)$ gravity'' 
consists of treating  the metric 
and the connection as independent variables ({\em i.e.},  the connection 
$\Gamma^{a}_{bc}$  is not the metric connection 
$\lbrace^{a}_{bc}\rbrace$ of 
$g_{ab}$).  If the (non-metric) connection is assumed to be 
symmetric, the field 
equations (which are now of second order) are \cite{Vollick}
\begin{eqnarray}
&& f^\prime\left( \tilde{R} \right) 
R_{ab}-\frac{f\left( \tilde{R} \right) }{2}g_{ab}=\kappa \, 
T_{ab}^{(m)} \;,\\
&&\nonumber \\
&& -\tilde{\nabla}_c \left[ \sqrt{-g} 
f'(\tilde{R}) 
g^{ab}\right] +\tilde{\nabla}_d \left[ \sqrt{-g} f'(\tilde{R}) 
g^{d(a}\right]\delta^{b)}_c  =0 \;,
 \end{eqnarray}
where $\tilde{R}_{ab}$ is the  Ricci tensor of the non-metric 
connection,  $\tilde{R}=g^{ab}\tilde{R}_{ab}$, and 
$\tilde{\nabla}_c $ is the covariant 
derivative operator of $\Gamma^a_{bc}$.

The third possibility, ``metric-affine gravity'', allows also 
the matter action to depend on the independent connection, 
resulting in more general field  equations 
\cite{metricaffine}. In this paper,  we only consider the metric 
and  Palatini formalisms.

First, let us focus on metric $f(R)$ gravity. In order for 
these theories
 to be viable, several criteria must be met, which have been analysed in 
the recent past. The theory must have the correct Newtonian and post-Newtonian limits 
\cite{CSE, weakfield} and  it must not have short timescale   
instabilities, 
which is achieved if  $f^{\prime\prime}>0$  
\cite{DolgovKawasaki, mattmodgrav, 
Odintsovconfirm, SawickiHu} (see Refs.~\cite{instabilities} for 
studies of 
various instabilities). 
The theory must possess the correct cosmological dynamics, {\em i.e.}, an inflationary 
era,  
followed by a radiation era and then a matter era and, finally, an acceleration era. 
Furthermore, there must be smooth transitions  between consecutive eras, which is not 
always achieved, with many models being ruled out   
\cite{Amendolaetal}.  
The theory must 
also be ghost-free, which is the case for $f(R)$ and for Gauss-Bonnet gravity, but not for  
theories of the form $f\left( R, R_{ab}R^{ab}, R_{abcd}R^{abcd}\right)$  
\cite{ghosts}.   
Another criterion for validity that should certainly be satisfied is that the theory  
possesses a well-posed  initial value formulation. Here we focus 
on that 
aspect. In addition to being a viability criterion, the Cauchy problem proves 
very useful in the numerical integration of the field equations.

The  initial value problem was studied for theories of the form
\begin{equation}
S = \int d^{4}x\sqrt{-g}\left(\frac{R}{2\kappa} +\alpha R_{ab}R^{ab}+\beta  
R^{2}+\gamma 
R_{abcd}R^{abcd}\right)+S^{matter}
\end{equation}
in Refs.~\cite{Noakes, TeyssandierTourrenc} with the conclusion 
that the Cauchy 
 problem is well-posed. Due to the Gauss-Bonnet identity in four spacetime  dimensions, 
the Kretschmann scalar can be dropped from the action. In the following, we  consider the 
Cauchy problem for the theory described by the action~(\ref{2}) and for a general form of 
the  function 
$f(R)$, without restricting to the specific theory $
f(R) = R + \alpha R^{2}$  considered in Refs.~\cite{Noakes,  
TeyssandierTourrenc}.  Instead 
of proceeding directly on the field equations (\ref{fieldeqs}), we will use a 
well known dynamical equivalence relation mapping $f(R)$ into  
scalar-tensor theories  \cite{STequivalence, 
TeyssandierTourrenc} 
and we will  address the Cauchy problem in  scalar-tensor 
gravity, 
relying on recent results on this subject  \cite{Salgado}. Our 
main result 
 is that the Cauchy problem for metric $f(R)$ gravity is well-formulated, 
 and is well-posed in vacuum. For  Palatini $f(R)$ gravity, 
instead, the  Cauchy problem is 
not well-formulated nor well-posed due to the presence of higher derivatives of  the 
matter  fields in the field equations and to the impossibility 
of eliminating 
them.  Here, the system of $3+1$ equations of motion is 
said to  be {\em well-formulated} if it can be re-written as a 
system of  only first-order equations (in time and space) in the 
scalar field variables. When this set can be put in 
the full  first order form 
\begin{equation}
\partial_t \vec{u} + M^i \nabla_i \vec{u}=\vec{S}\left( 
\vec{u}\right) \;,
\end{equation}
where $\vec{u}$ collectively denotes the fundamental variables 
$h_{ij}, K_{ij}$, {\em etc.} introduced below, $M^i$ is 
called the {\em characteristic matrix} of the system, and 
$\vec{S}\left(  \vec{u} \right)$ describes source terms and 
contains only the fundamental variables but not their 
derivatives. Then, the initial 
value formulation is {\em well-posed} if the system of partial 
differential equations is {\em symmetric hyperbolic} ({\em 
i.e.}, $M^i$ are symmetric)  and {\em strongly hyperbolic} ({\em 
i.e.}, $ s_iM^i$ has a real set of eigenvalues and a complete 
set of eigenvectors for  any 1-form $s_i$, and obeys some 
boundedness conditions). We refer the reader to 
Ref.~\cite{Solin} for more precise definitions and for a 
complete discussion.

The plan of this paper is the following: in Sec.~2 
 we recall the equivalence between $f(R)$ and scalar-tensor 
gravity. In  Sec.~3 we proceed 
to show that the Cauchy problem is well-formulated for the 
scalar-tensor equivalent  of 
metric $f(R)$ gravity. The scalar-tensor equivalent of  Palatini  
$f(R)$ gravity is 
considered in Sec.~4, and Sec.~5 contains a discussion and the conclusions.

\section{The equivalence between $f(R)$ and scalar-tensor gravity}
\setcounter{equation}{0}

We briefly review the equivalence between $f(R)$ and scalar-tensor gravity,  starting with 
the metric formalism. By introducing the extra  scalar field 
$\phi$, the modified gravity  
action
\begin{equation}
\label{6}
S = \frac{1}{2\kappa}\int d^{4}x\sqrt{-g} \, f(R) + S^{(matter)}
\end{equation}
can be rewritten as \cite{STequivalence, TeyssandierTourrenc}
\begin{equation}
\label{7}
S = \frac{1}{2\kappa}\int d^{4}x\sqrt{-g}\left[ \psi(\phi)R - V(\phi)  \right] + 
S^{(matter)}
\end{equation}
when $f^{\prime\prime}(R) \neq 0$, where
\begin{eqnarray}
\label{8}
\psi = f^\prime(\phi) \;, &
\,\,\,\,\,\,\,\,\,\,\,\,\,\,\;\;\;  V(\phi) = \phi f^\prime(\phi) - f(\phi) 
\;.
\end{eqnarray}
The action (\ref{7}) coincides with (\ref{6}) if $\phi = R$. Vice-versa,  
the variation of 
(\ref{7}) with respect to $\phi$ yields
\begin{equation}
R \, \frac{d\psi}{d\phi}-\frac{dV}{d\phi} = 
\left(R-\phi\right)f^{\prime\prime}(R) = 0 \;,
\end{equation}
which has no dynamical content but implies\footnote{An action  of the 
form~(\ref{6}) with 
the property 
that $f^{\prime\prime}(R) \neq 0$ is sometimes called ``$R$-regular''  
\cite{MagnanoSokolowski}.} that 
 $\phi = R$ 
when $f^{\prime\prime}(R)  
\neq 0$.
(\ref{7}) is a Brans-Dicke action of the form \cite{BransDicke} 
\begin{equation}
S = \frac{1}{2\kappa}\int d^{4}x\sqrt{-g}\left[ \psi R - \frac{\omega}{2} 
\nabla^{c}\psi\nabla_{c}\psi - U(\psi)\right] + S^{(matter)}
\end{equation}
with Brans-Dicke field $\psi$, $U(\psi) = V\left[ \phi(\psi) \right]$, and 
Brans-Dicke parameter 
$\omega = 0$. This is called O'Hanlon theory or ``massive dilaton gravity''  
and was originally introduced in order to produce a Yukawa term in the 
Newtonian potential when taking the Newtonian limit 
\cite{OHanlon}. The field equations are 
\begin{eqnarray}
&& G_{ab} = \frac{\kappa}{\psi} \, T_{ab}^{(m)} -\frac{1}{2\psi}\,  
U(\psi)g_{ab}+\frac{
1}{\psi}\left(\nabla_{a}\nabla_{b}\psi-g_{ab}\square\psi\right) \;,\\
&&\nonumber\\
&& 3\square\psi+2  U(\psi)-\psi\,\frac{dU}{d\psi}=\kappa \, T^{(m)} \;.
\end{eqnarray}
Let us consider now modified gravity in the  Palatini 
formalism. The Palatini action
\begin{equation}
S = \frac{1}{2\kappa}\int d^{4}x\sqrt{-g}\, f(\tilde{R}) + 
S^{(matter)} 
\end{equation}
is equivalent to 
\begin{equation}
\label{14}
S = \frac{1}{2\kappa}\int d^{4}x\sqrt{-g}\left[ 
f(\chi)+f^\prime(\chi)\left 
(\tilde{R}-\chi\right)\right] + S^{(matter)} \;,
\end{equation}
the variation of which with respect to $\chi$ yields  $ \chi = 
\tilde{R}$. 
One now uses the field $\phi = f^\prime(\chi)$ and takes 
advantage of the  fact that the 
curvature $\tilde{R}$ can be seen as the Ricci curvature 
 associated to the new metric 
\begin{equation}
h_{ab}=f^\prime(\tilde{R})g_{ab}
\end{equation}
conformal  to $g_{ab}$ \cite{Ferrarisetal, metricaffine}. The 
two conformal frames are physically equivalent (see the 
recent extensive discussion of \cite{FaraoniNadeau} and  
references therein). The transformation 
property of  the Ricci scalar under conformal transformations is 
\cite{Synge,Wald}
\begin{equation}
\tilde{R}=R+\frac{3}{2\phi}\nabla^{c}\phi\nabla_{c}\phi-\frac{3}{2}\square\phi \;,
\end{equation}
which allows the action (\ref{14}) to be rewritten as
\begin{equation}
\label{17}
S = \frac{1}{2\kappa}\int d^{4}x\sqrt{-g}\left[ \phi R + \frac{3}{2\phi} 
\nabla^{c}\phi\nabla_{c}\phi - V(\phi)\right] + S^{(matter)} \;,
\end{equation}
where an irrelevant boundary term has been omitted and
\begin{equation}
V(\phi)=\phi\chi(\phi)-f\left[ \chi(\phi)\right] \;.
\end{equation}
The action (\ref{17}) describes a Brans-Dicke theory with parameter $\omega= 
-3/2$. This theory is generally regarded as a pathological case 
\cite{mybook},  but 
is sometimes studied \cite{omegapathological}. 

A general Brans-Dicke theory
\begin{equation}
S_{BD} = \frac{1}{2\kappa}\int d^{4}x\sqrt{-g}\left[ \phi R - \frac{\omega}{\phi} 
\nabla^{c}\phi\nabla_{c}\phi - V(\phi)\right] + S^{(matter)}
\end{equation}
leads to an effective gravitational coupling appearing in a Cavendish 
experiment  (and in the 
theory of cosmological perturbations \cite{Boisseauetal})
\begin{equation}
G_{eff} = \frac{2\left(\omega+2\right)}{\left(2\omega+3\right)}\frac{1}{\phi}\;,
\end{equation}
which becomes ill defined for $\omega = -3/2$. Similarly, the  equation of 
motion 
for $\phi$,
\begin{equation}
\left(2\omega + 3 \right)\square\phi = \kappa \, T^{(m)} + 
\phi\frac{dV}{d\phi} 
- 2V(\phi)
\end{equation}
reduces to an algebraic identity when $\omega = -3/2$ and the  field $\phi$ 
becomes non-dynamical. This has obviously consequences for the   
initial value problem of 
the theory, which are discussed later.

Our strategy will be to study the Cauchy problem of $f(R)$ gravity by 
reducing it to the  initial value problem of $\omega = 0 ,  
-3/2$  Brans-Dicke theory. Nevertheless, we start by considering 
more  general scalar-tensor 
theories in the next section.

\section{The initial value problem of scalar-tensor gravity}
\setcounter{equation}{0}

The Cauchy problem of Brans-Dicke and  scalar-tensor gravity 
theories was  considered in 
Refs.~\cite{Noakes, TeyssandierTourrenc, CockeCohen}. 
Noakes \cite{Noakes} showed that the Cauchy problem is well-posed for a 
non-minimally coupled  scalar field $\phi$ with action
\begin{equation}\label{NMC}
S = \int d^{4}x\sqrt{-g}\left[ \left(\frac{1}{2\kappa} - \xi\phi^{2}\right)  R - 
\frac{1}{2}\nabla^{c}\phi\nabla_{c}\phi - V(\phi)\right] 
\end{equation}
without matter. Originally, this theory was introduced by 
regarding $\phi $ as a  form of 
(quantum) matter on a fixed curved background, not as a 
gravitational  scalar field akin to the Brans-Dicke field, nor 
as a source of  
gravity on 
the right hand side of the Einstein equations. There 
is little doubt that this is the reason why Noakes did not include ``ordinary'' matter in 
the theory, which would complicate the discussion and weaken his conclusion of 
well-posedness  of the Cauchy problem \footnote{This example 
shows that there is no clear-cut distinction  between 
``gravitational'' and  
``non-gravitational''  scalar fields in relativistic theories of 
gravity (see  
Ref.~\cite{SotiriouFaraoniLiberati} for a discussion and 
implications).}. 
However, (\ref{NMC}) can be legitimately regarded as a  
scalar-tensor action.

Cocke  and Cohen \cite{CockeCohen} sketched a study using 
Gaussian normal  coordinates for Brans-Dicke theory with a free 
Brans-Dicke  scalar. A covariant and systematic approach to the 
Cauchy problem of  scalar-tensor theories of the form
\begin{equation}
S = \int d^{4}x\sqrt{-g}\left[ \frac{f(\phi) R}{2\kappa}  - 
\frac{1}{2}\nabla^{c}\phi\nabla_{c}\phi  - 
V(\phi)\right] + S^{(matter)}
\end{equation}
has recently been  advanced by Salgado  \cite{Salgado}, who 
showed that the 
initial value  problem for these theories is well-posed in the 
absence of matter and  well-formulated 
otherwise. For our purposes it is necessary to generalize Salgado's results 
to more general  
scalar-tensor theories of the form
\begin{equation}
S = \int d^{4}x\sqrt{-g}\left[ \frac{f(\phi) R}{2\kappa} - \frac{\omega
(\phi)}{2}\nabla^{c}\phi\nabla_{c}\phi - V(\phi)\right] + S^{(matter)}\;,
\end{equation}
which contain the additional coupling function $\omega(\phi)$. In the following, 
we show in detail how this generalization proceeds and then specialize 
the discussion to the cases $\omega = 0,  -3/2 
$ which are the main focus of our interest. We follow as closely as possible 
the notations of Ref. \cite{Salgado} in order to facilitate comparison.

The field equations are (we set $\kappa = 1$ in the rest  
of this paper)
\begin{eqnarray}
&& G_{ab} = \frac{1}{f}\left[ f^{\prime\prime} \left( 
\nabla_{a}\phi\nabla_{b}\phi  - 
g_{ab}\nabla^{c}\phi\nabla_{c}\phi \right) + 
f^\prime \left( \nabla_{a}\nabla_{b} \phi - 
g_{ab}\square\phi \right)\right] \nonumber \\
&&\nonumber \\
 && + \frac{1}{f}\left[ \omega \left( 
\nabla_{a}\phi\nabla_{b}\phi - \frac{1}{
2} g_{ab}\nabla^{c}\phi\nabla_{c}\phi \right) - V(\phi)g_{ab} + 
T_{ab}^{(m)}
\right] \;,\label{24}\\
&&\nonumber \\
&& \omega\square\phi + \frac{f^\prime}{2}  R -  
V^\prime (\phi) +
\frac{\omega^\prime}{2}\nabla^{c} 
\phi\nabla_{c}\phi = 0 \;,
\end{eqnarray}
where a prime denotes differentiation with respect to $\phi$. Eq.~(\ref{24}) is rewritten 
as the effective Einstein equation
\begin{equation}
\label{26}
G_{ab} = T_{ab}^{(eff)} \;,
\end{equation}
where the effective stress-energy tensor is \cite{Salgado}
\begin{equation}
T_{ab}^{(eff)} = \frac{1}{f(\phi)}\left( T_{ab}^{\left( f \right)} + T_{ab}^{\left(  \phi  
\right)} + T_{ab}^{(m)}\right) \;,
\end{equation}
with
\begin{equation}
T_{ab}^{\left( f \right)} = f^{\prime\prime}(\phi) \left( 
\nabla_{a} \phi \nabla_{b} \phi - 
g_{ab}\nabla^{c}\phi\nabla_{c}\phi \right) + f^\prime (\phi)\left( 
\nabla_{a}\nabla_{b} \phi - 
g_{ab}\square\phi \right) \;,
\end{equation}
which contains second order derivatives of $\phi$ and is identical to the one 
in 
Ref.~\cite{Salgado}, and
\begin{equation}
T_{ab}^{(\phi)} = \omega(\phi) \left( \nabla_{a}\phi\nabla_{b}\phi - 
\frac{1}{2}g_{ab}\nabla^{c}\phi\nabla_{c}\phi \right) - V(\phi)g_{ab}
\end{equation}
which has canonical structure ({\em i.e.}, quadratic in the  
first-order derivatives of 
 $\phi$ with no second order derivatives) and differs from the  
one of Ref. 
\cite{Salgado}  by 
the presence of the coupling function $\omega(\phi)$. By taking 
the  trace of 
eq.~(\ref{26}) and solving for $\square\phi$, one obtains 
\begin{equation}
\square\phi = \frac{ \frac{f^\prime T^{(m)}}{2} -2f^\prime 
V(\phi) + f V^\prime (\phi) 
+ \left[  -\frac{\omega^\prime f}{2} -  \frac{f^\prime}{2} 
\left( \omega + 
3f^{\prime\prime} \right) \right] \nabla^{c}\phi\nabla_{c}\phi }{ f\left[ 
\omega+\frac{3\left(f^\prime\right)^{2}}{2f}\right] }\;.
\end{equation}
The $3+1$ ADM formulation of the theory proceeds by defining the  
usual quantities lapse, shift,  extrinsic curvature, and 
gradients of $\phi$ 
\cite{Wald, Reula,  Salgado}. Assume that a time function $t$ 
exists 
such that the spacetime ($M$,$g_{ab}$) can be foliated by a family of
 hypersurfaces $\Sigma_{t}$ of constant $t$ with unit timelike normal $n^{a}$.  The 
3-metric is defined by $h_{ab} = g_{ab}+n_{a}n_{b}$ and ${h^{a}}_{c}$
 is the projection operator on $\Sigma_{t}$, with
\begin{eqnarray}
n^{a}n_{a} = -1 \;, &\,\,\,\,\,\,\,\,\,\,\,\,\,\,\,\, h_{ab}n^{b} = h_{ab
}n^{a} = 0 \;, &\,\,\,\,\,\,\,\,\,\,\,\,\,\,\,\, {h_{a}}^{b} h_{bc} = h_{ac} \;.
\end{eqnarray}
The metric is written using the lapse function $N$ and the shift vector $
N^{a}$ as
\begin{equation}
ds^{2} = -\left( N^{2} - N^{i}N_{i} \right) dt^{2} - 2N_{i}dtdx^{i} + h
_{ij}dx^{i}dx^{j}
\end{equation}
$\left( i,j = 1,2,3 \right)$, with $N > 0$, $n_{a} = - N \nabla_{a} t
$ and 
\begin{equation}
N^{a} = -{h^{a}}_{b}t^{b} \;,
\end{equation}
where $t^{a}$ is a time flow vector satisfying $t^{a}\nabla_{a}t = 1$ and
\begin{equation}
t^{a} = -N^{a}+Nn^{a} 
\end{equation}
so that $N = -n_{a}t^{a}$ and $N^{a}n_{a} = 0$. The extrinsic curvature 
 of $\Sigma_{t}$ is
\begin{equation}
K_{ab} = -{h_{a}}^{c} {h_{b}}^{d}\nabla_{c}n_{d} 
\end{equation}
and the 3D covariant derivative of $h_{ab}$ on $\Sigma_{t}$ is defined by 
\begin{equation}
D_{i} ^{\left(3\right)}{T^{a_{1}\ldots}}_{b_{1}\ldots} = { h^{a_{1}}  }_{c_{1}}  \ldots 
{h^{d_{1}}}_{b_{1}}\ldots {h^{f}}_{i}\nabla_{ f} ^{ 
\left(3\right)} {T^{c_{1}\ldots}}_{d_{1}\ldots}
\end{equation}
for any 3-tensor $^{(3)} {T^{a_{1} \ldots} }_{b_{1}\ldots}$ , with $D_i h_{ab}= 0$.  The 
 spatial gradient of the scalar field is
\begin{equation}
Q_a \equiv  D_a \phi \;,
\end{equation}
while its momentum is
\begin{equation}
\Pi = {\cal L}_n\phi = n^c\nabla_c\phi 
\end{equation}
with
\begin{equation}
K_{ij} = -\nabla_i n_j = -\frac{1}{2N}\left(\frac{\partial h_{ij}}{\partial t}  + 
D_i N_j 
+ D_j N_i \right) \;,
\end{equation}
\begin{equation} 
\Pi = \frac{1}{N}\left(\partial_t\phi+N^cQ_c\right) \;,
\end{equation}
and
\begin{equation}
\partial_tQ_i+N^l\partial_lQ_i+Q_l\partial_iN^l=D_i\left(N\Pi\right) \;.
\end{equation}
The $3+1$  decomposition of the effective stress-energy tensor 
$T_{ab}^{(eff)}$  follows: 
\begin{equation}
T_{ab}^{(eff)} = \frac{1}{f}\left( S_{ab}+J_an_b+J_bn_a+En_an_b \right) \;,
\end{equation}
where
\begin{equation} \label{41}
S_{ab} \equiv {h_a}^c {h_b}^d T_{cd}^{(eff)} = 
\frac{1}{f}\left( S_{ab}^{(f)}+S_{ab}^{(\phi)} 
+S_{ab}^{(m)}\right) \;,
\end{equation}
\begin{equation}
\label{42}
J_{a} \equiv -{h_a}^c T_{cd}^{(eff)} n^d = \frac{1}{f}\left(J_{a}^{(f)}+J_{a}^{(
\phi)}+J_{a}^{(m)}\right) \;,
\end{equation}
\begin{equation}\label{43}
E \equiv n^a n^b T_{ab}^{(eff)} = 
\frac{1}{f}\left(E^{(f)}+E^{(\phi)}+E^{(m)}\right) \;,
\end{equation}
and $T^{(eff)} = S-E$, where $T^{(eff)} 
\equiv {{T^{(eff)}}^a}_{a}$ and $S \equiv {S^a}_a$.
By means of the Gauss-Codacci equations, the effective Einstein equations 
 projected tangentially and orthogonally to $\Sigma_t$ provide the Hamiltonian constraint 
 \cite{Wald, Salgado}
\begin{equation}
^{(3)}R + K^2 - K_{ij}K^{ij} = 2E \;,
\end{equation}
the vector (or momentum) constraint 
\begin{equation}
\label{45}
D_l {K^l}_i - D_i K = J_i \;,
\end{equation}
and the dynamical equations
\begin{eqnarray}
\label{46}
&& \partial_t {K^i}_j + N^l \partial_l {K^i}_j + {K^i}_l \partial_j N^l - 
{K^l}_j 
\partial_l N^i + D^i D_j N \nonumber \\
&&\nonumber \\
&&- ^{(3)}{R^i}_j N - NK{K^i}_j = \frac{N}{2}\left[ \left(S-E\right) 
\delta^i_j -2S^i_j 
\right] \;,
\end{eqnarray}
where $K \equiv {K^i}_i$. The trace of this equation yields
\begin{equation}
\partial_t K + N^l \partial_l K + ^{(3)}\Delta N - NK_{ij}K^{ij} = \frac{N}{2} \left( S + 
E \right) \;,
\end{equation}
where $^{(3)}\Delta \equiv D^i D_i$. As remarked in 
\cite{Noakes, Salgado},  the presence of second order 
derivatives of the scalar $\phi$ (seen here as an effective form 
of ``matter'') could potentially render the Cauchy problem   
ill-formulated. However, this is not the case because these 
derivatives can be eliminated in most cases (with one notable 
exception explained  later) \cite{Salgado}.

The $f$- and $\phi$-quantities appearing in eqs.~(\ref{41})-(\ref{43}) are then calculated, 
yielding
\begin{equation}
E^{(f)} = f^\prime\left( D^c Q_c + K\Pi \right) + f^{\prime\prime}Q^2\;,
\end{equation}
\begin{equation}
J_a^{(f)} = -f^\prime\left( K_a^c Q_c +D_a\Pi \right) - f^{\prime\prime
}\Pi Q_a \;,
\end{equation}
\begin{equation}
S_{ab}^{(f)} = f^\prime\left( D_a Q_b + \Pi K_{ab} - h_{ab}\square\phi 
\right) - f^{\prime\prime} \left[ h_{ab}\left( Q^2 - \Pi^2 \right) - Q_a Q_b 
\right] \;,
\end{equation}
where $Q^2 \equiv Q^c Q_c$. Other useful quantities are
\begin{equation}
S^{(f)} = f^\prime \left( D_c Q^c + K\Pi - 3\square\phi \right) +  f^{\prime\prime}\left( 
3\Pi^2 - 2Q^2 \right) \;,
\end{equation}
\begin{equation}
S^{(f)}-E^{(f)} = -3f^\prime\square\phi - 3f^{\prime\prime}\left( Q^2 - 
 \Pi^2 \right) \;.
\end{equation}
All these $f$-quantities coincide with those of Ref.~\cite{Salgado}. However,  
the 
following $\phi$-quantities differ from the corresponding ones of 
\cite{Salgado} by the presence of terms in $\omega$ and $\omega^\prime$:
\begin{equation}
E^{(\phi)} = \frac{\omega}{2}\left( \Pi^2 + Q^2 \right) + V(\phi) \;,
\end{equation}
\begin{equation}
J_a^{(\phi)} = -\omega\Pi Q_a \;,
\end{equation}
\begin{equation}
S_{ab}^{(\phi)} = \omega Q_a Q_b - h_{ab}\left[ \frac{\omega}{2}\left( 
Q^2 - \Pi^2 \right) + V(\phi) \right] \;,
\end{equation}
while
\begin{equation}
S^{(\phi)} = \frac{\omega}{2}\left( 3\Pi^2 - Q^2 \right) - 
3V(\phi) 
\end{equation}
and
\begin{equation}
S^{(\phi)}-E^{(\phi)} = \omega\left( \Pi^2 - Q^2 \right) - 
4V(\phi) \;.
\end{equation}
The ``total'' quantities entering the right hand side  of the $3+1$ field 
equations are then
\begin{equation}
E = \frac{1}{f} \left[ f^\prime\left( D^c Q_c + K\Pi \right) + 
\left( 
f^{\prime\prime}  +  \frac{\omega}{2}\right)Q^2 + 
\frac{\omega}{2}\Pi^2 + V(\phi) + 
 E^{(m)} \right] \;,
\end{equation}

\begin{equation}
J_a = \frac{1}{f}\left[ -f^\prime\left( {K_a}^c Q_c + D_a \Pi \right) -  
\left( 
f^{\prime\prime} + \omega \right) \Pi Q_a + J_a^{(m)} \right] \;,
\end{equation}

\begin{eqnarray}
&& S_{ab} = \frac{1}{f}\left\{ f^\prime \left( D_a Q_b + \Pi K_{ab} \right) 
 -h_{ab}\left[ \left( f^{\prime\prime}+\frac{\omega}{2} \right) \left( Q^2 - 
\Pi^2 \right) + V(\phi)  + f^\prime\square\phi \right]\right\} 
\nonumber \\
&&\nonumber \\
&& + \frac{1}{f}\left[ \left( \omega + f^{\prime\prime} \right) Q_a Q_b + S_{ 
ab}^{(m)} \right] \;,
\end{eqnarray}
while
\begin{eqnarray} 
&& S = \frac{1}{f}\left[ f^{\prime} \left( D^c Q_c + \Pi K 
\right) - 3V(\phi) - 
3f^\prime\square\phi \right] \nonumber \\
&&\nonumber \\
&&- \frac{1}{f}\left[ \left( 2f^{\prime\prime} + \frac{\omega}{2}\right)Q^2
 - 3\left( f^{\prime\prime} + \frac{\omega}{2} \right)\Pi^2 + S^{(m)}\right] 
\;,\\
&&\nonumber\\
&& S - E =  \nonumber \\ 
&&\nonumber\\
&& \frac{1}{f}\left[ \left(3f^{\prime\prime}+\omega\right) 
\left(\Pi^2-Q^2\right) - 4V(\phi) - 3f^\prime\square\phi + S^{(m)} - E^{(m
)}\right] \;,\\
&&\nonumber\\
&& S + E = \frac{1}{f}\left[ 2f^\prime\left( D^c Q_c + K\Pi \right) 
-f^{\prime \prime}Q^2 
+ \left( 3f^{\prime\prime} +2\omega \right)\Pi^2 \right] \nonumber \\
&&\nonumber \\
&& + \frac{1}{f}\left( - 2V(\phi)  - 3f^\prime\square\phi + 
S^{(m)}+ E^{(m)}\right) 
\;.
\end{eqnarray}
The Hamiltonian constraint becomes
\begin{eqnarray}
\label{64}
&& ^{(3)}R + K^2 - K_{ij}K^{ij} - \frac{2}{f}\left[ f^\prime\left( D_c Q^c + 
 K\Pi \right) + \frac{\omega}{2}\Pi^2 + \frac{Q^2}{2}\left( \omega + 2f^{
\prime\prime} \right) \right] \nonumber \\
&&\nonumber \\
&& = \frac{2}{f}\left( E^{(m)}+V(\phi) \right) \;,
\end{eqnarray}
the momentum constraint (\ref{45}) is 
\begin{equation} 
\label{65} 
D_l {K^l}_i - D_i K + \frac{1}{f}\left[ f^\prime \left( {K_i}^c Q_c + D_i\Pi 
\right) + \left( \omega + f^{\prime\prime} \right) \Pi Q_i \right] = 
\frac{J_i^{(m)}}{f} \;, 
\end{equation}
the dynamical equation~(\ref{46}) is written as
\begin{eqnarray}
&& \partial_t {K^i}_j + N^l \partial_l {K^i}_j + {K^i}_l 
\partial_j N^l - {K_j}^l  \partial_l N^i 
+ D^i D_j N - ^{(3)}{R^i}_j N - NK{K^i}_j \nonumber \\
&&\nonumber \\
&& + \frac{N}{2f}\left[ f^{\prime\prime}\left( Q^2 - \Pi^2 
\right) + 2V(\phi) +  f^\prime\square\phi  \right] 
\delta^i_j + \frac{Nf^\prime}{f}\left( D^i Q_j 
+ \Pi {K^i}_j \right) \nonumber \\
&&\nonumber \\
&&+ \frac{N}{f}\left( \omega + f^{\prime\prime} \right) Q^i Q_j 
= \frac{N}{2f} \left[\left( S^{(m)} - E^{(m)} \right) \delta^i_j 
- 2{S^{(m) \,\, i}}_j \right] \;,
\end{eqnarray}
with trace
\begin{eqnarray}
\label{67}
&& \partial_t K + N^l \partial_l K + ^{(3)}\Delta N - NK_{ij}K^{ij} - \frac{
Nf^\prime}{f}\left( D^c Q_c + \Pi K \right) \nonumber \\
&&\nonumber \\
&& + \frac{N}{2f}\left[ f^{\prime\prime}Q^2 - \left( 2\omega + 
3f^{\prime\prime}  
\right)\Pi^2 \right] = \frac{N}{2f}\left( -2V(\phi)  
-3f^\prime\square\phi + 
S^{(m)} + 
E^{(m)} \right) 
\end{eqnarray}
where (cf. Ref.~\cite{Salgado})
\begin{eqnarray}
\label{68}
&& {\cal L}_n\Pi - \Pi K - Q^c D_c \left( \ln N \right) - D_c 
Q^c  = -\square\phi
 \nonumber \\
&& =  -\frac{1}{f\left[ \omega + \frac{3{(f^\prime)}^2}{2f} 
\right]}\left(  \frac{f^\prime T^{(m)}}{2} -2f^\prime 
V(\phi) + fV^\prime (\phi) \right) \nonumber \\
&&\nonumber \\
&& -\frac{1}{f \left[ \omega + \frac{3{(f^\prime)}^2}{2f} 
\right]}\left\{ \left[
\frac{-\omega^\prime f}{2} - \left( \omega + 3f^{\prime\prime} 
\right)
\frac{f^\prime}{2} \right] \nabla^c\phi\nabla_c\phi \right\} \;.
\end{eqnarray}
In vacuo, the  initial data $(h_{ij}, K_{ij}, \phi, Q_i, \Pi)$ 
on an initial 
hypersurface $\Sigma_0$ must satisfy the  constraints~(\ref{64}) and (\ref{65}) plus 
\begin{equation}
Q_i - D_i\phi = 0 \;,
\end{equation}
\begin{equation}
D_i Q_j = D_j Q_i \;. 
\end{equation}
In the presence of matter, the variables $E^{(m)}$, $J_a^{(m)}$, $S_{ab}^{(m)}$ must  also 
be specified on the  initial hypersurface. Gauge-fixing is 
equivalent to  
assigning lapse 
$N$ and shift $N^a$ (various gauge conditions used in numerical analysis are reviewed in 
\cite{Salgado}). The system (\ref{64})-(\ref{67}) contains only  
first-order  derivatives 
in both space and time once the d'Alembertian $\square\phi$ is written in terms of 
$\phi, \nabla^c\phi\nabla_c\phi$, $f$, and its derivatives by using eq.~(\ref{68}).  This 
system differs from the corresponding one in \cite{Salgado} only by terms in  
$\omega(\phi)$ and its first derivative. From now on, everything will proceed as in 
Ref.~\cite{Salgado} (with one exception discussed  in the next section). The reduction to 
a first-order system  indicates that the Cauchy problem is 
well-posed in vacuo, 
this time also for the more general  scalar-tensor theories 
containing the extra 
coupling function $\omega (\phi)$, and that it is well-formulated 
 in the presence of matter. We do not repeat Salgado's analysis here, referring the 
reader to \cite{Salgado}. We are now ready to translate the results in terms of 
$f(R)$ gravity.

\section{The Cauchy problem of modified gravity}
\setcounter{equation}{0}

In the notations of Ref.~\cite{Salgado} that we adopted, 
Brans-Dicke theory  \cite{BransDicke} is described by $\omega 
(\phi) = \omega_0/  \phi $  with $\omega_0$ the constant 
Brans-Dicke parameter,  $f (\phi)
 = \phi$, and $V\rightarrow 2V$. This yields the constraints
\begin{eqnarray}
&& ^{(3)}R + K^2 - K_{ij}K^{ij} - \frac{2}{\phi}\left[ D^c Q_c + K\Pi +  
\frac{\omega_0}{2\phi}\left( \Pi^2 + Q^2 \right) \right] \nonumber \\
&&\nonumber \\
&&= \frac{2}{\phi}\left[ E^{(m)} + V(\phi) \right] \;, \label{71}\\
&&\nonumber\\
&& D_l {K^l}_i - D_i K + \frac{1}{\phi}\left( {K_i}^l Q_l + D_i \Pi + 
\frac{\omega_0}{ \phi}\Pi 
Q_i \right) = \frac{J_i^{(m)}}{\phi} \;,
\end{eqnarray}
and the dynamical equations
\begin{eqnarray}
&& \partial_t {K^i}_j + N^l\partial_l {K^i}_j + {K^i}_l\partial_j N^l - 
{K_j}^l\partial_l  N^i 
+ D^i D_j N \nonumber \\
&&\nonumber \\
&& - ^{(3)}{R^i}_j N - NK{K^i}_j + \frac{N}{2\phi}\delta^i_j 
\left( 2V(\phi) + 
\square
\phi \right) + \frac{N}{\phi} \left( D^i Q_j + \Pi {K^i}_j  \right) \nonumber 
\\
&&\nonumber \\
&& + \frac{N\omega_0}{\phi^2}Q^i Q_j = \frac{N}{2\phi}\left( \left( S^{(m)
} - E^{(m)} \right) \delta^i_j - 2 {S^{(m) \,\, i}}_j \right) \;,\\
&&\nonumber \\
&& \partial_t K + N^l\partial_l K + ^{(3)}\Delta N - NK_{ij}K^{ij} - \frac{N
}{\phi}\left( D^{c}Q_{c} + \Pi K \right) - \frac{\omega_{0}N}{\phi^2}\Pi^2 \nonumber \\
&&\nonumber \\
 &&= \frac{N}{2\phi}\left[ -2V(\phi) - 3\square\phi + S^{(m)} + E^{(m)} 
\right] \;,\label{74}
\end{eqnarray}
with
\begin{equation}
\label{75}
\left( \omega_0 + \frac{3}{2} \right)\square\phi = \frac{T^{(m)}}{2} - 
2V(\phi) + \phi V^{\prime}(\phi) + \mathbf{\frac{\omega_{0}}{\phi}}\left( \Pi^{2}
 - Q^{2} \right) \;. 
\end{equation}
>From the discussion of the previous sections, we conclude that metric $f
(R)$ gravity, which is equivalent to $\omega_0 = 0$ Brans-Dicke gravity,  has 
a well-formulated Cauchy problem in general and is also well-posed in vacuo.  On the 
other hand,  Palatini $f(R)$ gravity, which 
is equivalent to a Brans-Dicke theory with $\omega_0 = -3/2$, is in trouble. 
In 
fact, for this value of the Brans-Dicke parameter, the d'Alembertian $ \square\phi$ 
disappears from eq.~(\ref{75}) and the field  $\phi$ is not dynamical---it can be 
arbitrarily assigned on a region or on the entire spacetime,  provided its gradient 
satisfies the degenerate
 equation~(\ref{75}), which reduces to a constraint. 

As a consequence, in the  Palatini version, it is impossible to 
eliminate 
$\square\phi$ from the 
system~(\ref{71})-(\ref{74}) unless $\square\phi = 0$. This condition includes  both the 
possibility $\phi =$ constant, in which case the theory 
degenerates to general relativity, and the case of a harmonic $\phi$, which  is associated 
to a null $\phi$-wave. Apart from these special cases, the previous considerations 
constitute a no-go theorem for  Palatini $f(R)$ gravity,  which 
has 
an ill-formulated Cauchy problem in vacuo and, therefore, can hardly be regarded as a 
viable theory. The absence of a well-posed Cauchy problem was briefly noticed 
in Ref.~\cite{Noakes}.

\section{Discussion and conclusions}
\setcounter{equation}{0}
 
While metric $f(R)$ gravity has a well-formulated initial 
value problem (and a well-posed  
one in vacuo),  Palatini modified gravity does not. In 
conjunction with recent criticism of the Palatini formalism 
(according to which the correct Palatini variation necessitates 
a Lagrange multiplier method which is usually neglected and, 
when taken into account, produces essentially the same theory as 
the metric version \cite{Kaloperetal}---see also 
\cite{Deser}), this fact
probably leads to the demise of  Palatini $f(R)$ gravity  as a 
legitimate model of the cosmic acceleration. However, something 
can be learned from  the situation of the Palatini version of 
modified gravity. It  is interesting  to speculate 
on the possible physical consequences of the failure of a 
theory to have a well-posed  initial value problem for the  
reasons discussed  in the  previous section. As seen above, the 
difficulty originates from the fact  that the  effective 
``matter'', {\em i.e.}, the scalar field $\phi$, is  described 
by a stress-energy tensor containing derivatives of $\phi$ of 
second order instead of first.  This non-canonical form makes it 
very difficult, if not impossible, to satisfy  the energy 
conditions \cite{Wald}.  This feature in itself should not 
necessarily be regarded as a curse or failure. However, while 
for the $\omega=0$ equivalent of metric $f(R)$ gravity the 
second derivatives contained in $\Box\phi$ can be eliminated, 
for the $\omega=-3/2$  equivalent of Palatini $f(R)$ gravity 
this is not possible and, when integrating twice to solve 
the field equations, the solution for the metric will depend on $\phi$, not  only on an 
integral of $\phi$ (for example through the usual Green function 
integral).  Then, the metric will be very sensitive to 
variations in  $\phi$ that are not smoothed out, or averaged, by 
an integral. This extreme  sensitivity, contrary to ordinary 
situations, may spoil the continuous dependence on initial  
conditions. Physically, this is the situation encountered in 
modelling stars in  Palatini $f(R)$ gravity. In Ref. \cite{BSM} 
it  is found that the dependence of the metric on higher order 
derivatives of  the matter  fields makes it so sensitive to 
small variations that it is impossible to build even a model of 
a polytropic star, while a polytropic gas is a perfectly reasonable  
form of  matter even in a Newtonian star. It is auspicable that 
this  dependence on higher order derivatives of the matter 
fields be avoided in future theories of gravity, for example, 
by including higher order terms  in $ R_{ab}R^{ab}$ in the 
action.


\ack{We thank A.A. Starobinsky and T.P. Sotiriou for 
stimulating discussions. This 
work was supported by a Bishop's University Research 
Grant and by the Natural Sciences and Engineering Research 
Council of Canada (NSERC).}

\vskip1truecm



\clearpage 

\Bibliography{99} 

\bibitem{SN}  Riess A G {\em et al.} 1998 {\em Astron. J.} 
{\bf 116} 1009; 1999 {\em Astron. J.} {\bf 118} 2668;
2001 {\em Astrophys. J.} {\bf 560} 49;
2004, {\em Astrophys. J.} {\bf 607} 665;
Perlmutter S {\em et al.} 1998 {\em Nature} {\bf 391} 51; 
1999 {\em Astrophys. J.} {\bf 517} 565;  
Tonry J L {\em et al.} 2003 {\em Astrophys. J.} {\bf 594} 1; 
Knop R {\em et al.} 2003 {\em Astrophys. J.} {\bf 598} 102; 
Barris B {\em et al.} 2004 {\em Astrophys. J.} {\bf 602} 571;
Riess A G {\em et al.} astro-ph/0611572

\bibitem{MG1} Capozziello S, Carloni S and  
Troisi A 2003 preprint astro-ph/0303041;  Carroll S M, Duvvuri V,  Trodden M and  
Turner M S 2004 {\em Phys. Rev. D} {\bf 70} 043528 

\bibitem{Wald}  Wald R M 1984 {\em General Relativity} 
(Chicago: Chicago  University Press)

\bibitem{Schmidt}  Schmidt H-J 2007, {\em Int. J. Geom. 
Meth. Mod. Phys.} {\bf 4} 209 (preprint gr-qc/0602017)

\bibitem{EddingtonBuch} Weyl H 1918 {\em Stz. Preuss. Akad. Wiss.} {\bf 1} 
465; 1950 {\em Space, Time, 
Matter} (Dover, New York);
Bach R 1921 {\em Math. Zeischfr.} {\bf 9} 110;
Lanczos C 1932 {\em Zeitshfr. Phys.} {\bf 73}, 
147;
Eddington A 1925 {\em The Theory of Relativity} 
(Springer, Berlin);
Schr\"{o}dinger E 1950 {\em Space-Time 
Structure} (CUP, Cambridge);
Buchdhal H 1948   
{\em Quart. J. Math. Oxford} {\bf 19} 150; 
1948 {\em Proc. Nat. Acad. Sci. USA} {\bf 34} 66; 
1951 {\em Acta Mathematica} {\bf 85} 63; 
1951 {\em J. London Math. Soc.} {\bf 26} 139; 
1951 {\bf 26} 150; 
1953 {\em Proc. Edimburgh Math. Soc.} {\bf 10} 16; 
1962 {\em Nuovo Cimento} {\bf 23} 141; 
1979 {\em Tensor} {\bf 21} 340; 
1970 {\em Proc. Cambr. Phil. Soc.} {\bf 68} 179;
1970 {\em Mon. Not. R. Astr. Soc.} {\bf 150} 1; 
1973 {\em Proc. Canbr. Phil. Soc.} {\bf 74} 145; 
1978 {\em J. Phys. A} {\bf 11} 871; 
1978 {\em Int. J. Theor. Phys.} {\bf 17} 149 

\bibitem{renorma}  Utiyama R. and DeWitt B 1962 {\em J. Math. Phys}  
{\bf 3}, 608;
Stelle K S 1977, {\em Phys. Rev. D} {\bf 16}, 953; 
Strominger A 1984, {\em Phys. Rev. D} {\bf 30}, 2257;
Buchbinder I L, Odintsov S D and  Shapiro I L 1992 {\em 
Effective Action in  Quantum Gravity} (IOP, Bristol);
Vilkovisky G 1992 {\em Class. Quantum Grav.} {\bf 9}, 985.  
See also Codello A and  Percacci R 2006 {\em Phys. Rev. Lett.} {\bf 
97} 221301  

\bibitem{Starobinsky}  Starobinsky A A 1980 {\em Phys. Lett. B} {\bf 
91} 99;
Mihic M B,  Morris M S and Suen W-M 1986 {\em Phys. Rev. D} {\bf 34} 2934;
Amendola L,  Capozziello S,  
Litterio M and  Occhionero F 1992 {\em Phys. Rev. D} {\bf 45}  417;  
Occhionero F and Amendola L 1994 {\em Phys. Rev. D} {\bf 50} 4846

\bibitem{NOD} Nojiri S and Odintsov S D 2007 {\em Int. J. Geom. 
Meth. Mod. Phys.} {\bf 4} 115

\bibitem{Vollick} Vollick D N 2003 {\em Phys. Rev. D} {\bf 68} 
063510

\bibitem{metricaffine} Sotiriou T P and Liberati S 2007 {\em
  Ann. Phys.} {\bf 322} 935;  Poplawski N J 2006 {\em Class. 
Quantum Grav.} {\bf 23} 2011;  {\em Class. Quantum Grav.} {\bf 
23} 4819

\bibitem{CSE}  Erickcek A L, Smith T L  and  Kamionkowski M 
2006 {\em Phys. Rev. D} {\bf 74}  121501(R);   Chiba T, Smith T 
L and Erickcek A L 2007 {\em Phys. Rev. D} {\bf 75} 124014 

\bibitem{weakfield}
Soussa M E and Woodard R P 2004 {\em Gen. 
Rel. Grav.} {\bf 36} 855;  
Dick R 2004 {\em Gen. Rel. Grav.} {\bf 36} 217;
Dominguez A E and Barraco D E 2004 {\em Phys. Rev. D} {\bf 70} 043505;
Easson D A 2004 {\em Int. J. Mod. Phys. A} {\bf 19} 5343;
Olmo G J 2005 {\em Phys. Rev. Lett.} {\bf 95} 261102;
2005 {\em Phys. Rev. D} {\bf 72} 083505;
preprint gr-qc/0505135; preprint gr-qc/0505136;
Navarro I and Van Acoleyen K 2005 {\em Phys. Lett. B} {\bf 622} 1;
Allemandi G, Francaviglia M,  Ruggiero M L  and Tartaglia A 2005  
{\em Gen. Rel. Grav.} {\bf 37} 1891;
Cembranos J A R 2006 {\em Phys. Rev. D} {\bf 73} 064029;
Capozziello S and  Troisi A 2005 {\em Phys. Rev. D} {\bf 72}  
044022;
Clifton T and Barrow J D 2005 {\em Phys. Rev. D} {\bf 72} 103005; 
Sotiriou T P 2006 {\em Gen. Rel. Grav.} {\bf 38} 1407;
Shao C-G, Cai R-G, Wang B and 
Su R-K 2006 {\em Phys. Lett. B} {\bf 633} 164;
Faraoni V 2006 {\em Phys. Rev. D} {\bf 74} 023529;
Capozziello S, Stabile A and Troisi A 2006 {\em Mod. Phys. 
Lett. A} {\bf 21} 2291; 2007 preprint arXiv:0708.0723; 2007 
preprint arXiv:0709.0891; 
Allemandi G and Ruggiero M L 2006  preprint astro-ph/0610661;
2006 preprint astro-ph/0619661; Allemandi G, Francaviglia M, 
Ruggiero M L and Tartaglia A 2005 {\em Gen. Rel. Grav.} {\bf 37} 
1891;  Jin X-H, Liu D-J, and Li X-Z 2006  
astro-ph/0610854;
Zakharov A F, Nucita A A, De 
Paolis F, and  Ingrosso G 2006 {\em Phys. Rev. D} {\bf 74} 107101

\bibitem{DolgovKawasaki} Dolgov A D and Kawasaki 2003 {\em Phys. Lett. B}  
{\bf 573}  1

\bibitem{mattmodgrav} Faraoni V 2006 {\em Phys. Rev. D} {\bf 74} 
104017 

\bibitem{Odintsovconfirm}  Nojiri S and Odintsov S D 2003 {\em 
Phys. Rev. D} {\bf 68} 123512; 2004 {\em Gen. Rel. Gravit.}  {\bf 36} 1765;
Nojiri S 2004 {\em TPU Vestnik} {\bf 44N7}: 49 
(preprint hep-th/0407099);  Multamaki T and Vilja I 2006 {\em Phys. Rev. D} 
{\bf 73} 024018 

\bibitem{SawickiHu} Sawicki I and  Hu W 2007 {\em Phys. Rev. 
D} {\bf 75} 127502;  
 Song Y-S,  Hu W  and Sawicki I 2007 {\em Phys. Rev. D} 
{\bf 75} 044004; Nojiri S and Odintsov S D 2007 {\em Phys. Lett. 
B} {\bf 652} 343; preprint arXiv:0707.1941


\bibitem{instabilities}  Navarro I and Van 
Acoleyen K 2006 {\em J. Cosmol. Astropart. Phys.} {\bf 03} 008; 
Cognola G,  Elizalde E,  Nojiri S, 
Odintsov S D and  Zerbini S 2005, {\em J. Cosmol. Astropart. 
Phys.} {\bf 
02} 010; 2006 {\em J. Phys. A} {\bf 39} 6245; 2006 {\em J. 
Phys. A} {\bf 39} 6245; 
Cognola G,  Gastaldi M and  Zerbini S 2007, preprint 
gr-qc/0701138;
Clifton T and  Barrow J D 2005 {\em Phys. Rev. D} {\bf 72} 
123003; 2006 {\em Class. Quantum 
Grav.} {\bf 23} L1; 2005 {\em Phys. Rev. D} {\bf 72} 103005;
N\'{u}nez A and Solganik S 2005 {\em Phys. Lett. B} 
{\bf 608} 189;  Chiba T 2005 {\em J. Cosmol. Astropart.  Phys.} 
{\bf 0503}  008;
Wang P 2005  {\em Phys. Rev. D} {\bf 72} 024030; Barrow J D and  
Hervik S 2006, {\em 
Phys. Rev. D} {\bf 73}, 023007 (2006);
Calcagni G,  de Carlos B and De Felice A 2006  
{\em Nucl. Phys. B} {\bf 752} 404;
Dolgov A and  Pelliccia D N 2006  
{\em Nucl. Phys. B} {\bf 734} 208;
Traschen J and  Hill C T 1986 {\em Phys. Rev. D} {\bf 33} 3519; 
M\"{u}ller  V, Schmidt H-J and Starobinsky A A 1988 {\em Phys. 
Lett.} {\bf 
202B} 198;   Schmidt H-J 1988 {\em Class. Quantum Grav.} 
{\bf 5} 233; Battaglia Mayer A and Schmidt H-J 1993 {\em Class. 
Quantum Grav.} {\bf 10} 2441;
Faraoni V 2007 {\em Phys. Rev. D} {\bf 75} 067302; 
2005 {\em Phys. Rev. D} {\bf 72} 061501(R); 
2004 {\em Phys. Rev. D} {\bf 70} 044037; 
Faraoni V and Nadeau S 2005 {\em Phys. Rev. D} {\bf 72} 124005;  
Setare M R 2007 {\em Phys. Lett. B} {\bf 644} 99;
Sotiriou T P 2007  {\em Phys. Lett. B} {\bf 645} 389;
Bertolami O 1987 {\em Phys. Lett. B} {\bf 186} 161;
Faraoni V 2001 {\em Int. J. Theor. Phys.} {\bf 40} 2259;
Briscese F, Elizalde, E, Nojiri S and Odintsov S D 2007 {\em 
Phys. Lett. B} {\bf 646} 105


\bibitem{Amendolaetal}  Amendola L, Polarski D 
and Tsujikawa S 2007, {\em Phys. Rev. Lett.} {\bf 98} 131302;
Capozziello S,  Nojiri S,   
Odintsov S D  and Troisi A 2006 {\em Phys. Lett. B} {\bf 639} 135 
;  Amendola L,  
Polarski D and  Tsujikawa S 2006 preprint  astro-ph/0605384;  
Nojiri S and  
Odintsov S D 2006 {\em Phys. Rev. D} {\bf 74} 086005; 2007 {\em 
J. Phys. A} {\bf 40} 6725; 2007 {\em J. Phys. Conf. Ser.} {\bf 
66} 012005;   
Amendola L, Gannouji R,  Polarski D, 
and Tsujikawa S 2007, {\em Phys. Rev. D} {\bf 75} 083504;
Brookfield A W,  
van de  Bruck C  and  Hall L M C 2006 {\em Phys. Rev. D} {\bf 74}  
064028;
Capozziello S,  Cardone V and Troisi A 2005 {\em Phys. Rev. D} {\bf 71}
 043503;
Cruz-Dombriz A and  Dobado A 2006 {\em Phys. Rev. D} {\bf 74}  
087501;
Fay S, Nesseris S and  Perivolaropoulos L 2007 {\em Phys. Rev. 
D} {\bf 76} 063504; 
Fay S,  Tavakol R and Tsujikawa S 2007 {\em Phys. Rev. D} {\bf 
75} 063509 

\bibitem{ghosts} Stelle K 1978 {\em Gen. Rel. Grav.} {\bf 9} 
353;
Ferraris M, Francaviglia M and Magnano G 1988 {\em Class. 
Quantum Grav.} {\bf 5} L95;
N\'{u}nez A and  Solganik S 2004 preprint  
hep-th/0403159; 2005 {\em Phys. Lett. B} {\bf 608} 189;
Chiba T 2005 {\em JCAP} 03: 008;
De Felice A, Hindmarsh M and   Trodden M 2006 {\em JCAP} 
0608:005;
Riegert R J 1984 {\em Phys. Lett. A} {\bf 
105} 110;
Navarro I and  Van 
Acoleyen K 2006 {\em J. Cosmol. Astropart. Phys.} {\bf 03} 008;
Comelli D 2005 {\em Phys. Rev. D} {\bf 72} 064018 

\bibitem{Noakes} Noakes D R 1983 {\em J. Math. Phys.} {\bf 24} 
1846

\bibitem{TeyssandierTourrenc} Teyssandier P and Tourrenc P 1983 
\JMP {\bf 24} 2793

\bibitem{STequivalence}  Higgs P W 1959 {\em Nuovo Cimento} 
{\bf 11} 816;   
Whitt B 1984 {\em Phys. Lett. B} {\bf 145}  176;  Wands D 1994 {\em 
Class. Quantum Grav.} {\bf 11} 269;
Chiba T 2003  {\em Phys. Lett. B} {\bf 575} 1

\bibitem{Salgado} Salgado M 2006 {\em Class. Quantum Grav.} {\bf 23} 4719

\bibitem{Solin} Sol\'{i}n P 2006 {\em Partial Differential 
Equations and the Finite Element Method} (New York: Wiley)

\bibitem{MagnanoSokolowski} Magnano G and Sokolowski L 1994 {\em 
Phys. Rev. D} {\bf 50} 5039 

\bibitem{BransDicke}  Brans C H and  Dicke R H  1961 {\em Phys. 
Rev.}  {\bf 124}  925 

\bibitem{OHanlon} O'Hanlon J 1972 {\em Phys. Rev. Lett.} {\bf 
29} 137

\bibitem{Ferrarisetal} 
Ferraris M, Francaviglia M and  Volovich I 1993, preprint 
gr-qc/9303007

\bibitem{FaraoniNadeau} Faraoni V and Nadeau S 2007 {\em Phys. 
Rev. D} {\bf 75} 023501

\bibitem{Synge}  Synge J L 1955, {\em Relativity: The General Theory}
(North Holland, Amsterdam)

\bibitem{mybook} Faraoni V 2004 {\em Cosmology in Scalar-Tensor 
Gravity} (Kluwer Academic, Dordrecht)

\bibitem{omegapathological}  Anderson J L 1971 {\em Phys. Rev. D} {\bf 
3} 1689; O'Hanlon J and  Tupper B 1972, {\em Nuovo Cimento 
B} {\bf 7} 305; O'Hanlon J 1972, {\em J. Phys. A} {\bf 5} 803; 
Deser S 1970 {\em Ann. Phys. (NY)} {\bf 59} 248; 
Barber G A 2003 preprint gr-qc/0302088;
Davidson A 2005 {\em Class. Quantum Grav.} {\bf 22} 1119;
Dabrowski M P, Denkiewicz T and Blaschke D 2007 {\em Ann. Phys. (Leipzig)} {\bf 
16}, 237 (hep-th/0507068)

\bibitem{Boisseauetal} Boisseau B, Esposito-Far\`{e}se G, 
Polarski D and Starobinsky A A 2000 {\em Phys. Rev. Lett.} 
{\bf 85} 2236

\bibitem{CockeCohen} Cocke J and Cohen J M 1968 {\em J. Math. 
Phys.} {\bf 9} 971

\bibitem{SotiriouFaraoniLiberati} Sotiriou T P, Faraoni V and 
Liberati S 2007 preprint arXiv:0707.2748

\bibitem{Reula} Reula O A 1998 {\em Living Rev. Rel.} {\bf 1} 3

\bibitem{BSM} Barausse E, Sotiriou T P and Miller J C 2007, 
preprint gr-qc/0703132

\bibitem{Kaloperetal} Iglesias A, Kaloper N, Padilla A and Park 
M 2007, preprint arXiv:0708.1163

\bibitem{Deser} Deser S 2006 {\em Class. Quantum Grav.} {\bf 23} 
5773

\endbib
\end{document}